\documentclass[a4paper]{jpconf}
\usepackage[english]{babel}
\usepackage{graphicx}
\usepackage{verbatim}
\usepackage{amsfonts}
\usepackage{amsmath}
\usepackage{makeidx}
\usepackage{color}
\begin{document}

\title{Jet measurements at STAR}


\author{Elena Bruna for the STAR Collaboration}

\address{INFN, Via P. Giuria 1, 10125 Torino, Italy}

\ead{bruna@to.infn.it}

\begin{abstract}
 Jets are produced from hard scatterings in the early stages of heavy-ion 
collisions, therefore they can be exploited as probes for medium tomography.
Such high-$p_T$ partons are expected to suffer energy loss in the hot and dense nuclear medium via gluon 
radiation or elastic collisions along their path.
Jet reconstruction gives access to the kinematics of the hard scattering that produced the jet, improving our understanding of energy loss and its effect on the jet structure.
Such measurements are challenging in heavy-ion collisions at RHIC, due to the large background, therefore a precise characterization of the background in Au+Au is needed.
We present an overview of the results on jet measurements obtained by the STAR experiment in p+p, d+Au and central Au+Au collisions at 200 GeV.
We also present results obtained with reconstructed di-jets and jet-hadron correlations as tools to study the medium effects on jet production.

 \end{abstract}



\subsection{Motivation and experimental techniques}
Hard probes produced in initial scattering in heavy-ion collisions are exploited to investigate the properties of the hot and dense medium, in particular to extract information on the mechanisms by which high-momentum quarks and gluons interact and lose energy.
Full jet reconstruction has been developed at RHIC to access the kinematics of the hard scattering and to quantitatively study the jet quenching phenomenon.
The STAR experiment is well suited for jet analysis because of the full azimuthal and the large rapidity coverage offered by both the Time Projection Chamber (TPC) for charged particles and the Barrel Electromagnetic Calorimeter (BEMC) for the neutral energy.
The jet-finding algorithms are based on recombination algorithms included in the FastJet package~\cite{fastjet1}. We use Anti-$k_T$ to reconstruct the signal jets and $k_T$ to estimate background per unit area.
The analyses reported in these proceedings are based on p+p year 2006, d+Au year 2008 and central Au+Au year 2007 events. 

\subsection{p+p and d+Au: the reference}

\begin{figure}
\vspace{-9mm}
\centering
\resizebox{0.49\textwidth}{!}{  \includegraphics{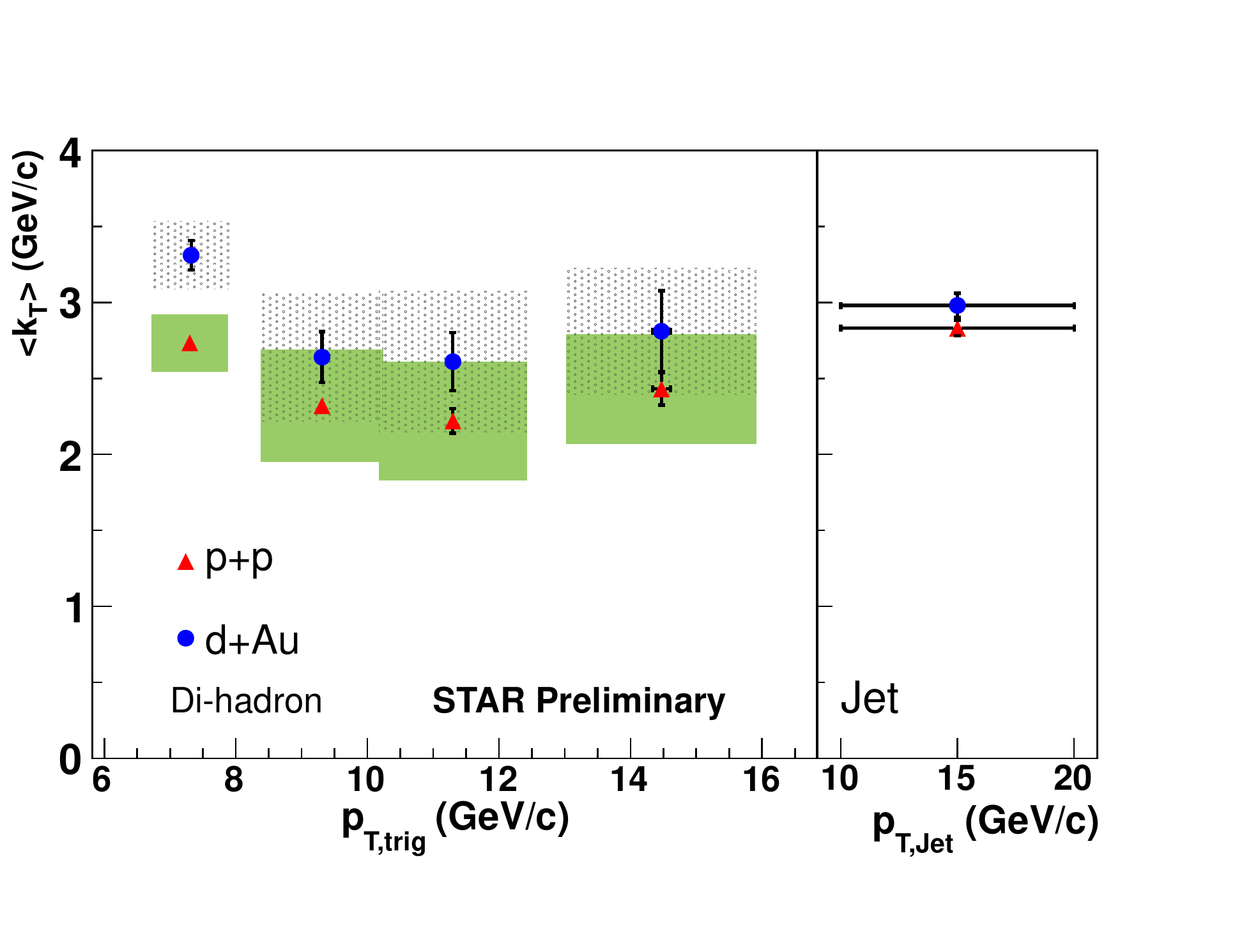}}
\resizebox{0.49\textwidth}{!}{  \includegraphics{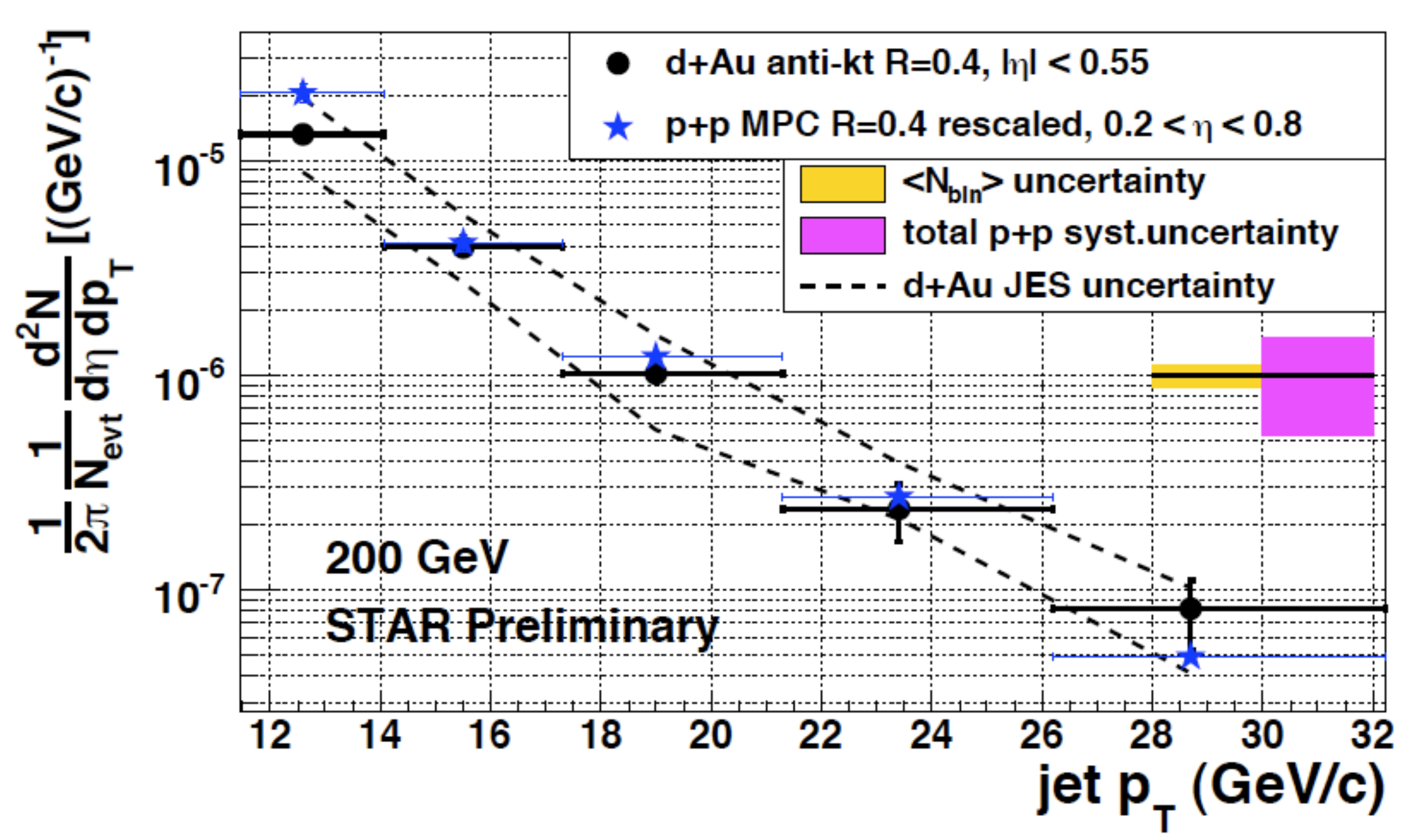}}
\vspace{-3mm}
\caption{Left: values of $\langle k_T\rangle$ for d+Au and p+p, measured from di-hadron and di-jet correlations. Vertical bars represent statistical errors, colored bands represent the systematic uncertainties, horizontal bars are the bin widths. Right: reconstructed jet $p_T$ spectrum in d+Au and binary-scaled p+p.}
\label{fig:kt}
\end {figure}

The measurement of jet cross section in p+p provides a baseline to assess nuclear effects in d+Au and Au+Au.
The STAR results on the cross section of fully reconstructed jets in p+p using the mid-point cone algorithm with cone radius 0.7 are reported in~\cite{tai}.
The data are compared to NLO calculations that take into account corrections for hadronization and the underlying event. The good agreement between data and theory indicates that reconstructed jets are calibrated probes in p+p to be utilized as reference for measurements in more complex systems, such as d+Au and Au+Au.
The performance of conceptually different jet-finding algorithms (i.e. SISCone, Anti-k$_T$ and k$_T$ from FastJet) were studied on the same p+p data set and it was found that the raw jet spectra agree within 10\%. 
Jet analysis on the d+Au data set allows to assess the effects of the cold nuclear matter, since no Quark Gluon Plasma is expected to be created. To this end, we look for an additional broadening (in d+Au relative to p+p) of the $\langle k_T\rangle$, representing the mean transverse momentum kick given to the di-jets. The Au nucleus might indeed affect the $\Delta\phi$ distribution of di-jets because nuclear effects, like scatterings of jet fragments in the d+Au medium, can occur. The  $\langle k_T\rangle$ value has been measured in STAR with both di-hadron and di-jet correlations~\cite{ktbroad,jan}. As shown in Fig.~\ref{fig:kt} (left), the measured $\langle k_T\rangle$ is systematically higher for d+Au compared to p+p, for all the measured p$_T$ ranges of the jet and trigger particles. This result suggests a small cold nuclear matter effect that manifests as deflection of the di-jet axis in the transverse plane.
A first indication that the cold nuclear matter does not significantly affect the measured jet yield in d+Au is given by the comparison of the d+Au jet spectrum with the same spectrum measured in p+p and binary scaled, shown in Fig.~\ref{fig:kt} (right)~\cite{jan}. The observed agreement of the two spectra within the experimental errors is expected for hard processes, in absence of nuclear effects. However, this is a preliminary study where Anti-k$_T$ was used for d+Au, whereas the p+p results were obtained with a midpoint cone algorithm.

\subsection{The underlying event}
\begin{figure}
\vspace{-9mm}
\centering
\resizebox{0.47\textwidth}{!}{  \includegraphics{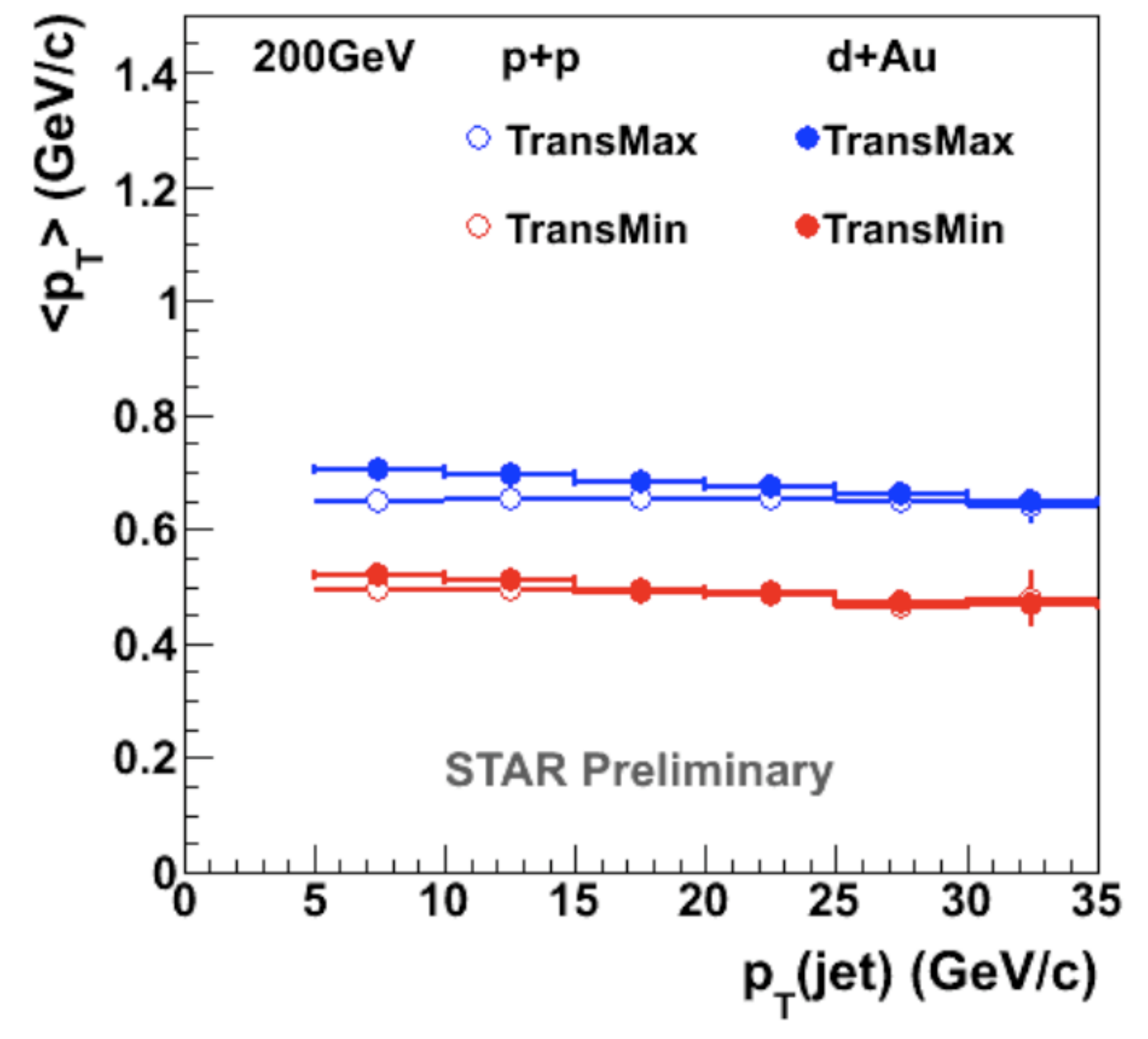}}
\resizebox{0.5\textwidth}{!}{  \includegraphics{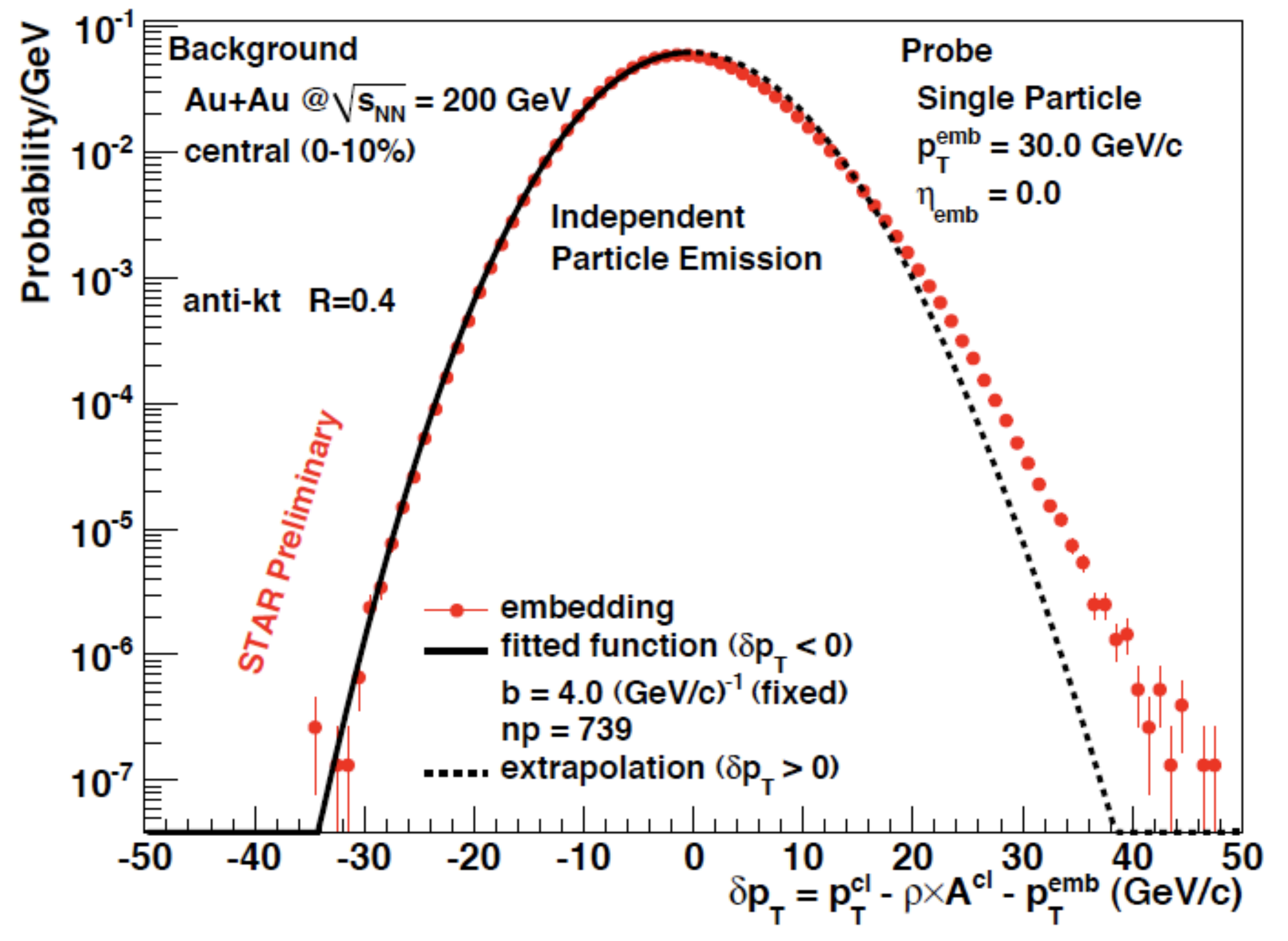}}
\vspace{-3mm}
\caption{Left:  $\langle p_T\rangle$ for d+Au and p+p, measured in the TransMax and TransMin regions relative to the di-jet axis, as a function of the leading jet $p_T$. Right: $\delta p_T$ distribution for single particle embedded probe. The LHS of the distribution is fitted with a Gamma function representing an independent particle emission model.}
\label{fig:ue}
\end {figure}

The underlying event consists of particles that are not correlated to the initial hard scatterings.
Following the approach used by CDF~\cite{cdf}, the underlying event in p+p and d+Au is investigated in the region transverse to the di-jet axis, where the jet contribution is expected to be minimal. In this region, two sub-sectors are define in the region of $60^\circ<\Delta \phi < 120^\circ$ and $-120^\circ<\Delta \phi < -60^\circ$ relative to the leading jet: the one with the largest charged particle multiplicity, called the TransMax region, and the other, called the TransMin region. The TransMax sector has a larger probability of
containing contributions from the hard initial and final state radiation of the scattered parton.
The mean transverse momentum measured in the TransMin and TransMax regions (Fig.~\ref{fig:ue}, left) is very similar in p+p and and d+Au and independent of the jet $p_T$, suggesting that the underlying event is decoupled from the hard scattering (~\cite{jana}).

The underlying event is the dominant source of particles in Au+Au collisions. The background density per unit area (on the $\eta-\phi$ plane), called $\rho$, is calculated as the median $p_T$ of all the clusters in a given event reconstructed with the k$_T$ algorithm.
Once the flat background is subtracted from the measured jet momentum, the jet spectrum is still affected by region-to-region fluctuations that distort the true jet momentum and are the source of ``fake jets''. Since the jet production is a hard process, we expect a scaling of the measured jet cross section in Au+Au relative to p+p, that can be expressed, taking into account the effect of the underlying event, as $\frac{d\sigma_{AA}}{dp_T} \propto [\frac{d\sigma_{pp}}{dp_T} +\rho A]\bigotimes F(A,p_T)$ , where $F(A,p_T)$ represents the background fluctuations under a jet of area $A$ and transverse momentum $p_T$.
A data-driven approach used in STAR to characterize the background fluctuations is the so called general probe embedding. In this method, a probe of a given momentum is embedded in a real Au+Au event. After running the jet-finding, the cluster containing the probe is found and $\delta p_T = p^{jet}_{T,rec} - \rho A - p_{T,probe}$ is calculated. Repeating the procedure for many probes over many events, one can assess the shape of the fluctuations under a ``signal'' probe over several orders of magnitude. Figure~\ref{fig:ue} (right) shows the $\delta p_T$ obtained by embedding single 30 GeV/c particles in the 0-10\% most central Au+Au events.  The left-hand side of the distribution ($\delta p_T <0$) is fitted with a Gamma function, expected to be describing the fluctuations in case of statistically independent particle emission~\cite{tannembaum}.
 The good agreement between data and fit on the LHS indicates that the bulk of the background fluctuations is accounted for by a model based on uncorrelated particle emission. As expected, the RHS deviates from the independent emission model because of  the presence of true jet signals.
 More systematic studies using different probes (jet signals generated with PYTHIA and qPYTHIA) besides the single particles indicate that background fluctuations do not have significant deviation when the fragmentation pattern is varied~\cite{jacobs}. This is crucial to characterize the background under quenched jets, whose fragmentation is a priori unknown.

\subsection{Jet broadening and softening in Au+Au}

Different complementary measurements based on full jet reconstruction seem to favor the hypothesis of broadening and softening of the jet energy profile in central Au+Au collisions.
In the jet-broadening scenario (possibly caused by large-angle gluon radiation) we expect the jet energy to be more diluted outside the cone radius compared to the reference p+p. This effect would result in a deficit of energy in the cone area when compared to p+p and a consequent shift of the jet $p_T$ spectrum towards lower $p_T$.
 
 One measurement that supports the existence of such an effect is the 
inclusive jet measurement in 0-20\% central Au+Au events in STAR~\cite{mateusz}.
Preliminary results show that the jet $R_{AA}$ is substantially larger than the single hadron $R_{AA}$ but still less than unity, suggesting that the jet algorithms fail to recover the partonic energy in a given radius.
In addition, the ratio of the yields of jets reconstructed with R=0.2 relative to R=0.4 is much less in central Au+Au compared to p+p. This indicates a deficit of jet energy within R=0.2 with respect to R=0.4, that can be explained as a spread of the jet profile to large areas caused by gluon radiation emitted at large angles.

\vspace{0.3cm}

Another measurement that is consistent with the conclusion of jet 
broadening and softening in Au+Au collisions is the di-jet measurement.
Di-jet measurements allow to study jet quenching in an extreme condition where the distance traversed by one of the two partons is maximized.
High momentum trigger jets, reconstructed with Anti-k$_T$ from particles above 2 GeV/c, are selected by the online trigger requiring one tower in the calorimeter to have $E_{tow}>5.4$ GeV, which implies a high-z fragmentation. Because of this selection, such trigger jets tend to be coming from partons that scatter near the surface of the medium. This surface bias is responsible for a larger, on average, in-medium path-length traversed by the recoil partons.
After selecting trigger jets with $p_T>20$ GeV in both p+p and in 0-20\% central Au+Au, we compared the per-trigger recoil jet spectrum (coincidence rate) in Au+Au relative to p+p. The preliminary results~\cite{elena} indicate a suppression of approximately a factor of 5, much stronger than the inclusive jet $R_{AA}$ and consistent with a path-length dependence of the energy loss.
\vspace{0.3cm}

\begin{figure}
\centering
\resizebox{0.49\textwidth}{!}{  \includegraphics{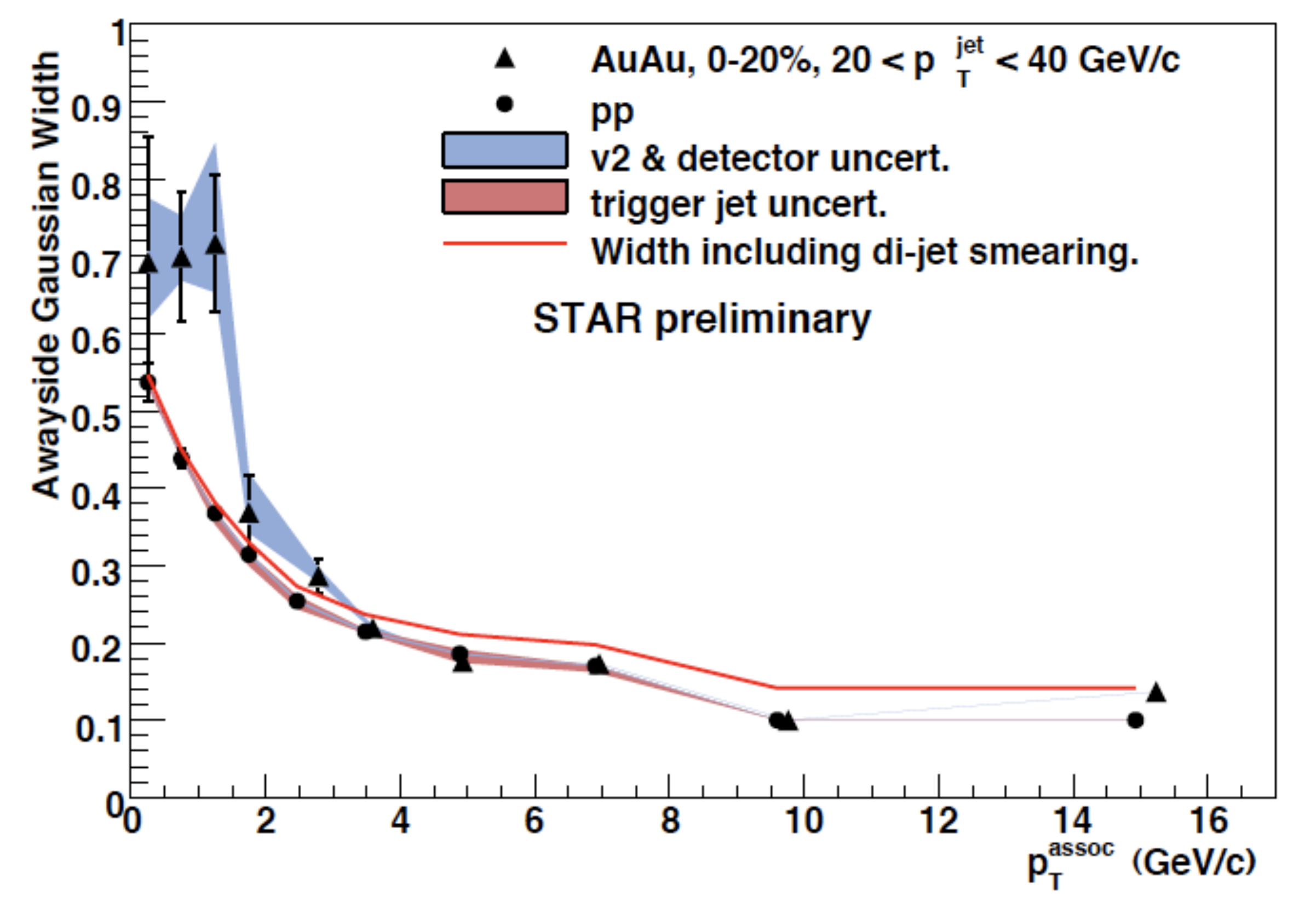}}
\resizebox{0.49\textwidth}{!}{  \includegraphics{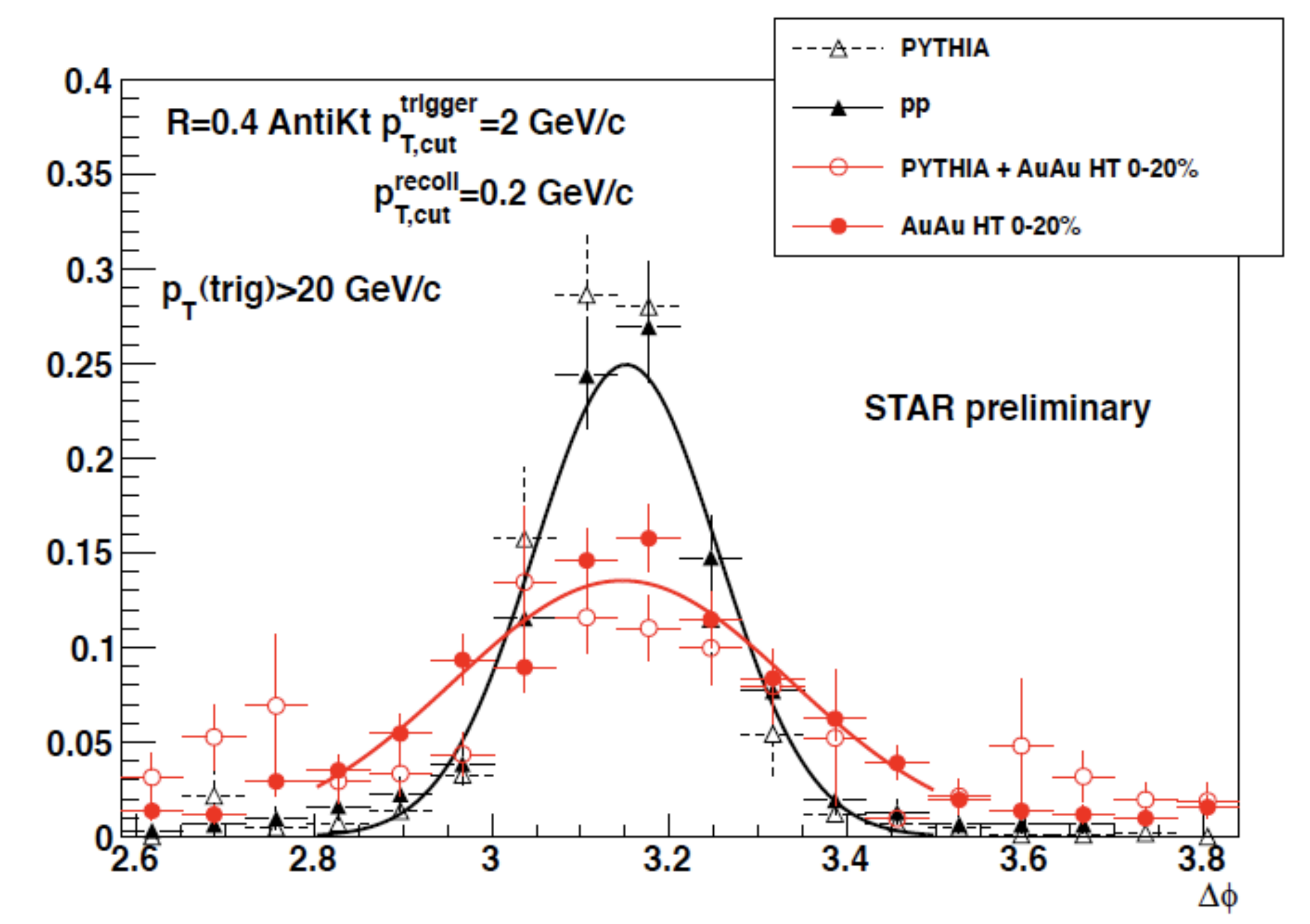}}
\caption{Left: width of the away-side jet-hadron correlations as a function of the associated $p_T$ of the charged particles, for p+p and Au+Au. Right:  $\Delta \phi$ distribution of reconstructed di-jets for PYTHIA, p+p, Au+Au and PYTHIA embedded in Au+Au events.}
\label{fig:JH}
\end {figure}

\begin{figure}
\centering
\resizebox{0.48\textwidth}{!}{  \includegraphics{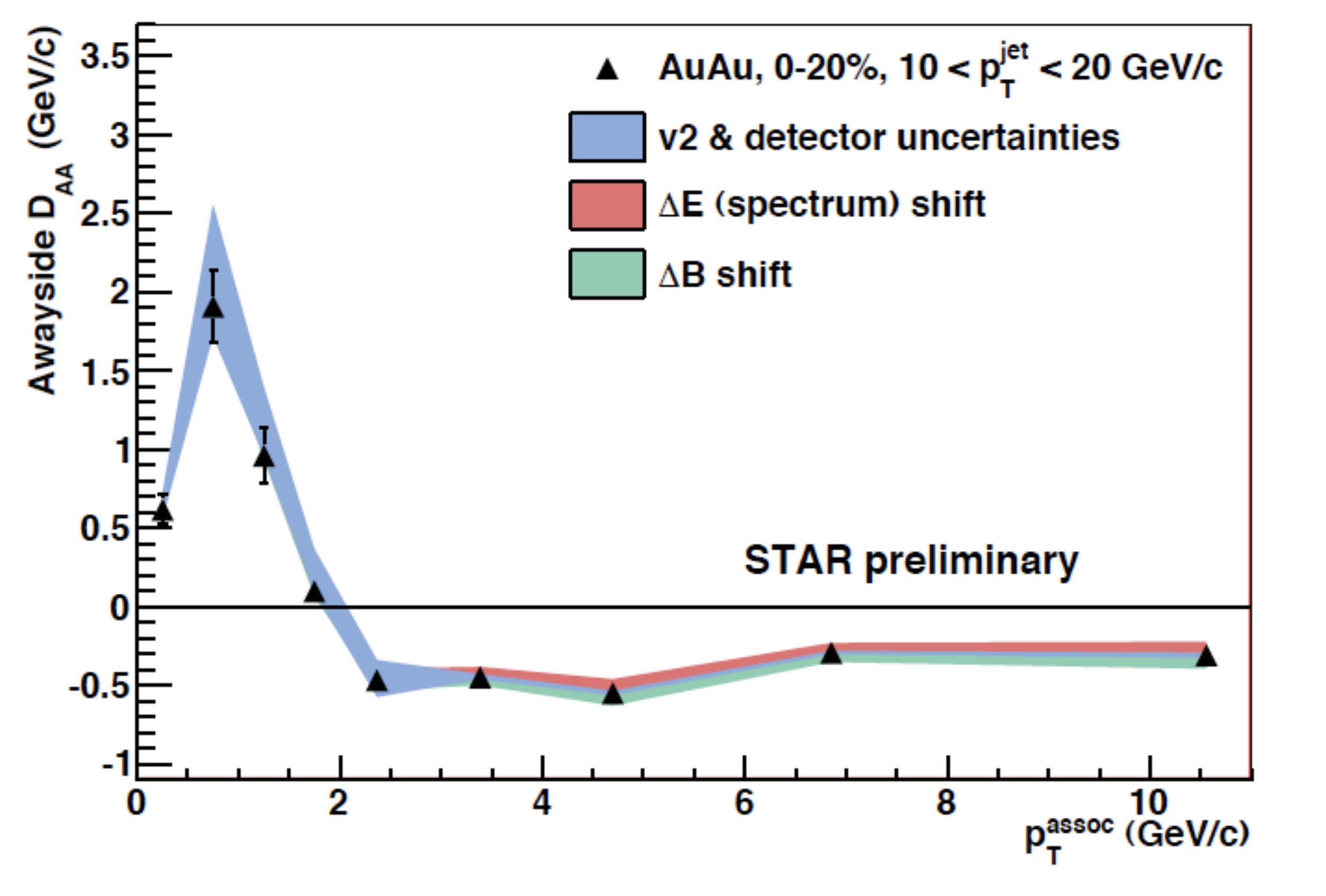}}
\resizebox{0.51\textwidth}{!}{  \includegraphics{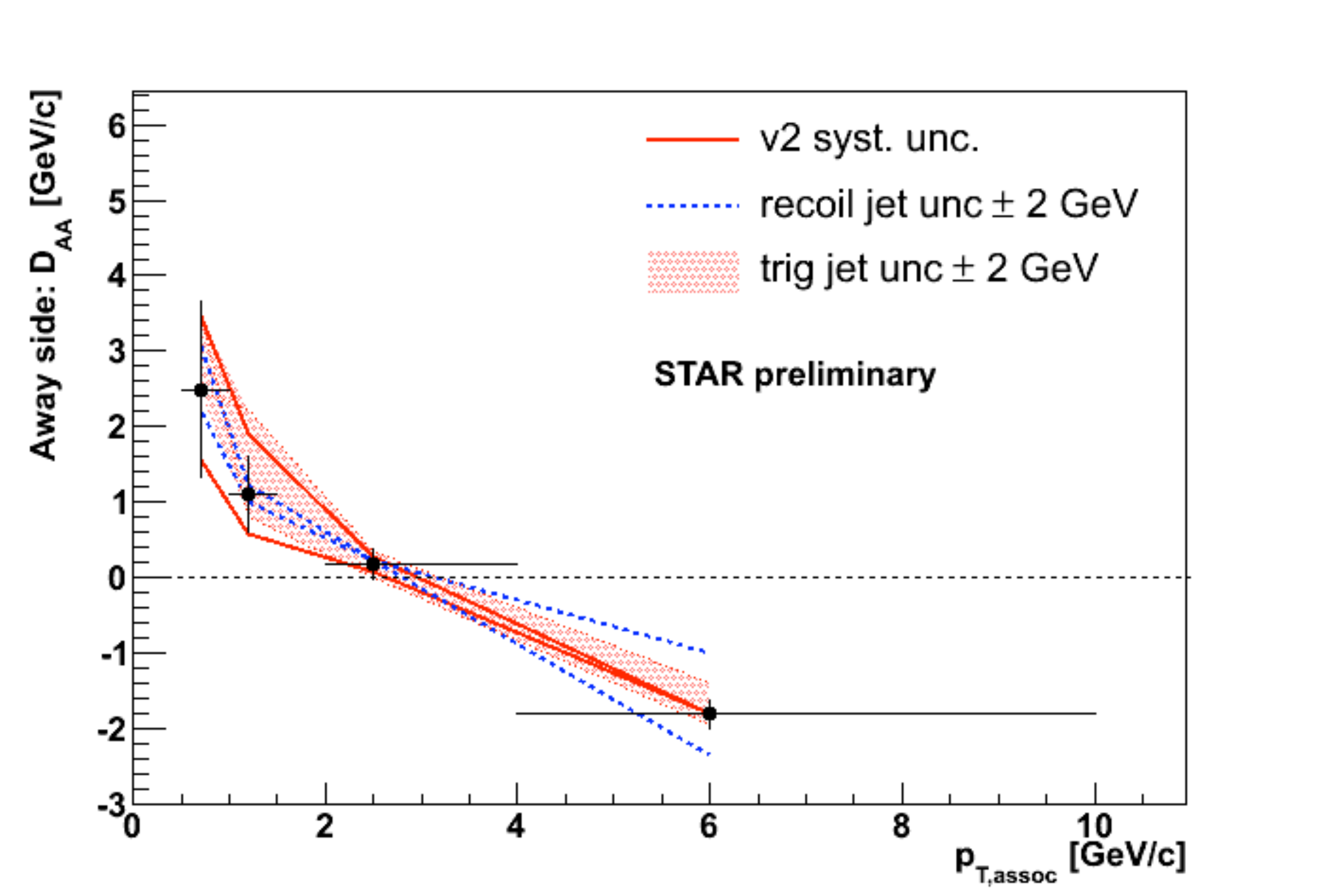}}
\caption{Left: $D_{AA}$ for the away-side of jet-hadron correlations as a function of the associated particle $p_T$. Right: same as left panel, but for events containing a reconstructed di-jet pair. Both trigger and recoil jets are reconstructed with R=0.4 and $p_{T,cut}=2$ GeV/c.}
\label{fig:JH2}
\end {figure}

To further investigate the mechanism of jet broadening we studied jet-hadron correlations~\cite{aliceO}. Once the trigger jet is selected (with the same criteria described above), azimuthal $\Delta \phi$ correlations relative to the jet axis are performed for the charged particles in the event.
Differently from di-hadron correlations, jet-hadron correlations allow a higher kinematical reach for the parton $p_T$, as well as a more precise measurement of the jet axis.
Figure~\ref{fig:JH} (left) shows the width of the away-side $\Delta \phi$ from jet-hadron correlations for p+p and Au+Au as a function of the $p_{T,assoc}$ of the charged particles entering the correlations. At low $p_{T,assoc}$ a broadening is observed in Au+Au compared to p+p. At higher  $p_{T,assoc}$, the width of the away side is similar in Au+Au and p+p, suggesting that hard fragmenting jets are biased toward the surface of the medium.
We looked at the $\Delta \phi$ distribution of fully reconstructed di-jets to evaluate the effect of the di-jet azimuthal smearing in jet-hadron correlations.
The $\Delta \phi$ distributions of di-jets reconstructed in PYTHIA, p+p, Au+Au and PYTHIA embedded in Au+Au events are reported in Fig.~\ref{fig:JH} (right). The rather small medium effect on the di-jet smearing is not sufficient to reproduce the measured broadening in jet-hadron correlations, as indicated by the red line in Fig.~\ref{fig:JH} (left).
Jet-hadron correlations also show a softening of the jet fragmentation, measured as an enhancement of low-$p_T$ particles in the away-side with a corresponding decrease of high-$p_T$ associated particles. This is quantitatively expressed by the difference of the integrated away-side yields in Au+Au and p+p, $D_{AA} = Y_{AA} \times  \langle p_{T,assoc} \rangle_{AA} - Y_{pp} \times  \langle p_{T,assoc} \rangle_{pp}$, as shown in Fig.~\ref{fig:JH2} (left).

\vspace{0.3cm}

A complementary analysis on jet-hadron correlations was performed in events with reconstructed di-jets (see also~\cite{elenaHP} for the details).
 As for the trigger jets, also recoil jets are reconstructed with R=0.4 and a $p_{T,cut}=2$ GeV/c on the particles to be used in the jet-finding. 
 Jet-hadron correlations can therefore show possible biases of the jet-finding algorithms. For instance, if the jet-finding biases the reconstructed di-jets towards being emitted tangentially with respect to the surface of the medium, we would expect similar away-side yields and widths in Au+Au and p+p for all the measured range of $p_{T,assoc}$. On the contrary, in spite of the strong bias given by the $p_{T,cut}$, the softening of the recoil jets is clearly visible in Fig.~\ref{fig:JH2} (right), where the $D_{AA}$ as a function of $p_{T,assoc}$ is shown for the events with reconstructed di-jet pairs.

\subsection{Summary}

Measurements of fully reconstructed jets in p+p collisions at STAR are in agreement with NLO calculations, providing a reference of well calibrated probes for d+Au and Au+Au measurements.
The jet analysis in d+Au was meant to quantify the effect of the cold nuclear matter on jet results. A rather small $\langle k_T \rangle$ broadening was seen, relative to the $\langle k_T \rangle$ measured in p+p. This does not affect the jet cross-section which is consistent, within the errors, with the binary-scaled p+p results.
The underlying event  in p+p and d+Au is decoupled from the hard scattering, with 
no substantial change in the $\langle p_T \rangle$.
A crucial aspect in the jet analysis is the treatment of the background in Au+Au and its fluctuations under the signal jets. 
There has been a major development in characterizing the background using a data-driven approach. 
Complementary jet measurements have been carried on in STAR: inclusive jet cross-section, di-jet coincidence rate and jet-hadron correlations. They  all converge toward a picture of broadening and softening of the jet fragmentation. This is in particular evident in the jet-hadron correlation results, where a broader width of the associated particles on the away-side is observed, together with an increase of soft particles and a corresponding decrease of the hard component. The observed broadening and softening  also for the highly biased population of di-jets suggests that the jet finding is not biased towards only surface or non-interacting jets.


\end{document}